\documentclass[final,1p,times]{elsarticle} 
\usepackage{graphicx}
\usepackage{amssymb} 
\usepackage{amsthm} 
\usepackage{lineno}

\newcommand{\be}{\begin{equation}}
\newcommand{\ee}{\end{equation}}
\newcommand{\ba}{\begin{eqnarray}}
\newcommand{\ea}{\end{eqnarray}}
\newcommand{\bd}{\begin{displaymath}}
\newcommand{\ed}{\end{displaymath}}

\journal{Nuclear Physics A} 

\begin{document}

\begin{frontmatter} 

% Your Title - please insert
\title{Influence of a Critical Point on Hydrodynamic Fluctuations in Heavy Ion Collisions}

%% Single author (and collaboration) - please insert
\author[auth1]{Joseph I. Kapusta$^{1,}$}
\fntext[col1] {Presenter.}
%\address{School of Physics \& Astronomy, University of Minnesota, Minneapolis, MN 55455, USA}

%% Multiple authors
\author[auth2]{Juan M. Torres-Rincon}
\address[auth1]{School of Physics \& Astronomy, University of Minnesota, Minneapolis, MN 55455, USA}
\address[auth2]{Departamento de Fisica Teorica I, Universidad Complutense de Madrid, 28040 Madrid, Spain}

\begin{abstract}
 
Hydrodynamic fluctuations have been studied in a wide variety of
physical, chemical, and biological phenomena in the past decade.  In high
energy heavy ion collisions, there will be intrinsic fluctuations even if the initial conditions
are fixed.   Fluctuations will be greatly enhanced if the trajectory in the plane of temperature versus chemical potential passes near a critical point.  We construct a model for the thermal conductivity which diverges at the critical point with the correct critical exponents, and use it in a simple illustrative model of a heavy ion collision.  The proton correlation function is sensitive to the presence of the critical point. 

\end{abstract} 

\end{frontmatter} % do not change

%% linenumbers are useful for reviewing process
%\linenumbers

The major reason to study QCD at finite temperature and density and to perform experiments with high energy heavy ion collisions is to map the phase structure of QCD in terms of temperature, baryon density, and isospin asymmetry.  The results have import for cosmology and for neutron stars, in addition to its intrinsic interest.  There are reasons to expect that there is a curve of first-order phase transition in the temperature $T$ versus baryon chemical potential $\mu$ plane, terminating at a critical point at $T_c > 0$ and $\mu_c > 0$ (with the simplifying condition of isospin symmetric matter).   The existence of such a critical point has been found in many effective field theory models.  Lattice QCD has yet to confirm or deny the existence of a critical point; for reviews, see Refs. \cite{stephanov} and \cite{MohantyQM}, and for up-to-date results see other papers in these proceedings.    

Experiments relevant to this inquiry involve low energy runs at the Relativistic Heavy Ion Collider (RHIC) and, in the future, at the Facility for Antiproton and Ion Research (FAIR), at the SPS Heavy Ion and Neutrino Experiment (SHINE), and at Nuclotron-based Ion Collider Facility (NICA).  Critical points are characterized by large fluctuations.  

A crucial issue is whether the critical point can ever be reached in a heavy ion collision.  Colliding nuclei is necessary to create high baryon density matter, but at the same time it creates entropy.  If the initial entropy per baryon is much larger than that at the critical point, then the expanding matter will never pass close to it, even under the assumption of an ideal adiabatic expansion: too much entropy is created.

A defining characteristic of a second-order phase transition, which in this case is the termination point of a line of first-order phase transition, is that fluctuations diverge.  These fluctuations and divergences have been studied in the context of QCD; see, for example, \cite{Kapusta2010}.  However, those studies were done under the conditions of a finite volume, static system in contact with a heat reservoir.  Those are not the conditions in a high energy heavy ion collision.  How can those fluctuations be quantified under the appropriate experimental conditions?

This question was addressed in \cite{Kapusta2012}; see also the paper by M. Stephanov in these proceedings.  It requires the use of hydrodynamic fluctuations, or noise, as discussed in \cite{statphys2}.  Stochastic terms are added to the energy-momentum tensor and to the conserved currents.  The average values of these stochastic terms are zero while the correlation functions are proportional to Dirac delta functions (white noise) with coefficients uniquely determined in terms of the dissipative coefficients, such as shear and bulk viscosity and thermal conductivity.  This is a direct consequence of the fluctuation-dissipation theorem.

In \cite{Kapusta2012} the focus was on the shear viscosity and heavy ion collisions at such high energy that the produced matter may be considered as having zero net baryon number.  In this work we focus on high baryon densities, hopefully achievable in lower energy heavy ion collisions, and neglect viscosity altogether.  The baryon current (in the Landau definition of flow velocity) is written as
\be
J^{\mu} = n u^{\mu} + \Delta J^{\mu} + I^{\mu} \ ,
\ee
where $n$ is the local baryon density, $u^{\mu}$ is the flow velocity, $\Delta J^{\mu}$ is the dissipative contribution, and $I^{\mu}$ is the noise term.  The latter has the correlation function
\be
\langle I^{\mu}(x_1) I^{\nu}(x_2) \rangle = 2 \lambda \left(\frac{nT}{w}\right)^2
\left(  u^{\mu} u^{\nu} - g^{\mu\nu} \right) \delta (x_1 - x_2) \, ,
\ee
where $\lambda$ is the thermal conductivity and $w$ is the enthalpy density.  The thermal conductivity diverges at a critical point which means that fluctuations are enhanced.  We use mode-coupling theory, which is very successful in describing the behavior of atomic and molecular systems near a critical point, combined with the study of \cite{Kapusta2010}, to parametrize the enhanced thermal conductivity (above the smooth background) as 
\be
\Delta \lambda = c_P \Delta D_T = c_P \frac{R_D T}{6\pi\eta_{\rm shear}\xi} \Omega(q_D\xi) \ .
\ee
\vspace{-0.32in}
\begin{figure*}[hb]
\begin{center}
\includegraphics[width=0.51\textwidth]{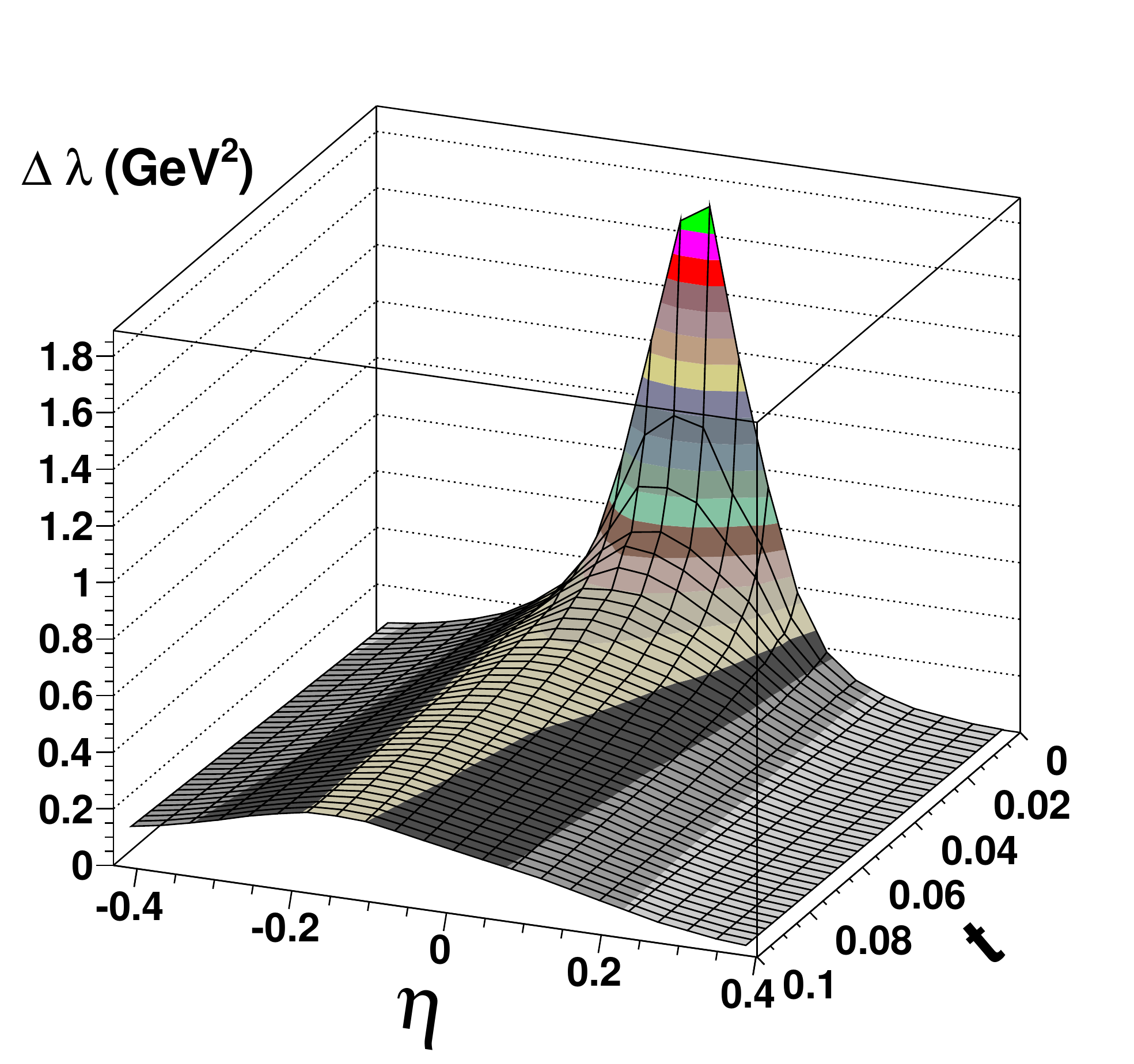}
\end{center}
\caption{Plot of $\Delta \lambda$ using the parametrization given in \cite{JoeJuan}. }
\label{fig:lambda_lego}
\end{figure*}
\newpage
Here $c_P$ is the isobaric heat capacity, $\xi$ is the correlation length, $\eta_{\rm shear}$ is the shear viscosity, $R_D$ is a universal constant approximately equal to 1.05, and $\Omega$ is a crossover function which goes to zero as $q_D \xi$ goes to zero and which goes to 1 as $q_D \xi$ goes to infinity. The $q_D$ is a cutoff in wavenumber and is material-dependent.  As the critical point is approached, $\xi \rightarrow \infty$, and the Stokes-Einstein diffusion law is recovered.  Our results for $\Delta \lambda$ in terms of the reduced temperature $t=(T-T_c)/T_c$ and reduced density $\eta=(n-n_c)/n_c$ are shown in Fig. 1.

To see whether fluctuation effects might have significant observable consequences we studied a simple illustrative model: the 1+1 dimensional boost invariant Bjorken model.  Although it is not a realistic model for describing heavy ion collisions that might create very high baryon density, it does allow for nearly analytic solution.  We solved the smooth perfect fluid equations of motion using the equation of state
\be
P(T,\mu)=A_4 T^4 + A_2 \mu^2 T^2 + A_0 \mu^4 - C T^2 - B
\ee
with coefficients $A_i$ appropriate to a plasma with 3 colors and 3 flavors of massless quarks; the coefficient $C$ was chosen to mimic the equation of state computed with lattice QCD at zero chemical potential.  Fluctuations were treated as perturbations and solved in the usual way using the method of Green functions.

In Fig. 2 the solid curve represents the line of first-order phase transition which terminates at a second-order phase transition - the critical point.  The dashed curve represents the crossover transition, as seen in lattice QCD calculations of the equation of state.  Also shown are three adiabatic trajectories, each one passing increasingly closer to the critical point.  The closer the trajectory comes to the critical point the more it feels the effects of the divergent thermal conductivity and therefore fluctuations.  These fluctuations give rise to correlations.  For example, if 
\begin{figure*}[hb]
\begin{center}
\includegraphics[width=0.61\textwidth]{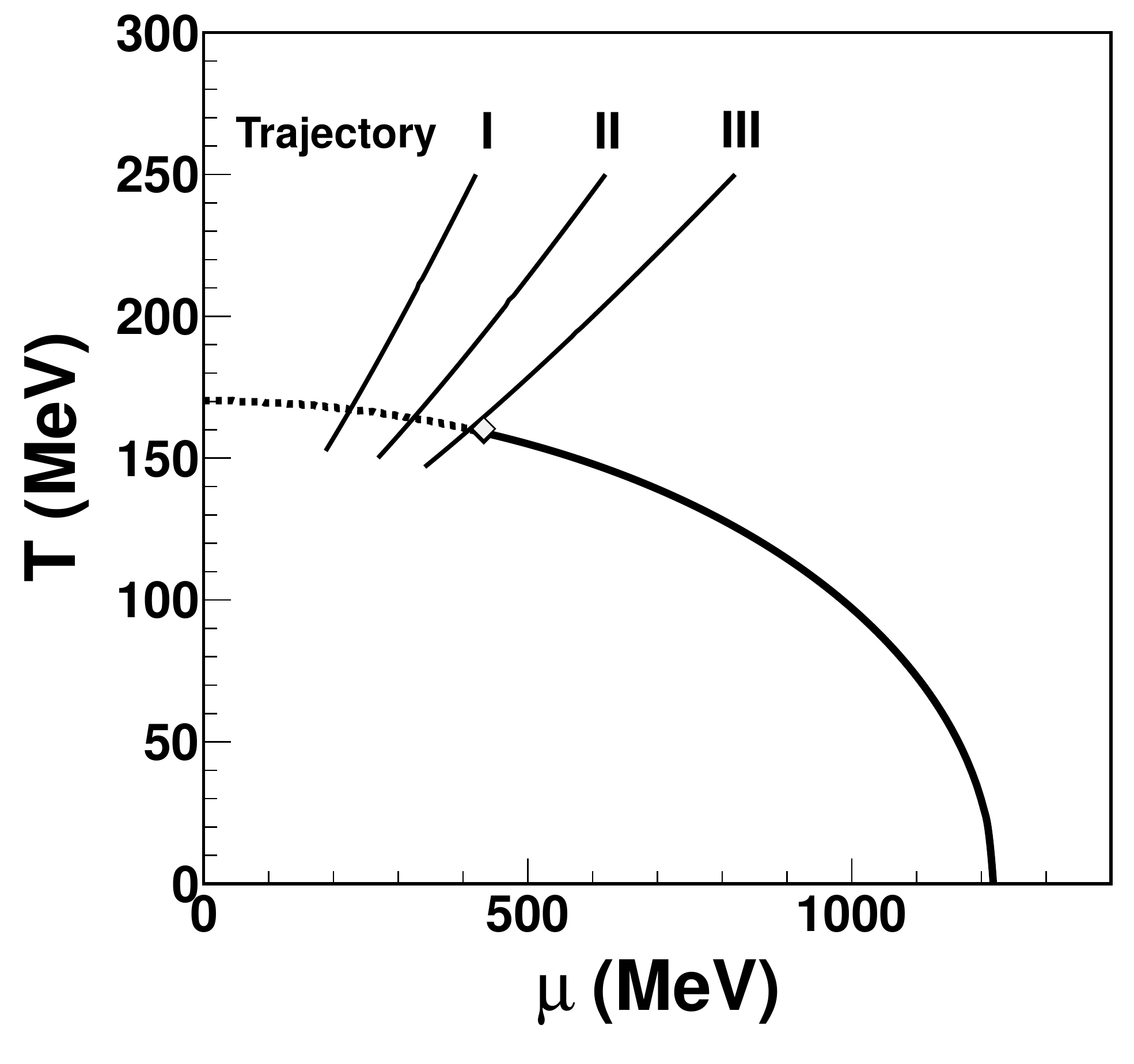}
\end{center}
\caption{Three adiabatic trajectories passing by the critical point.}
\label{fig:trajectories}
\end{figure*}
\begin{figure*}[ht]
\begin{center}
\includegraphics[width=0.75\textwidth]{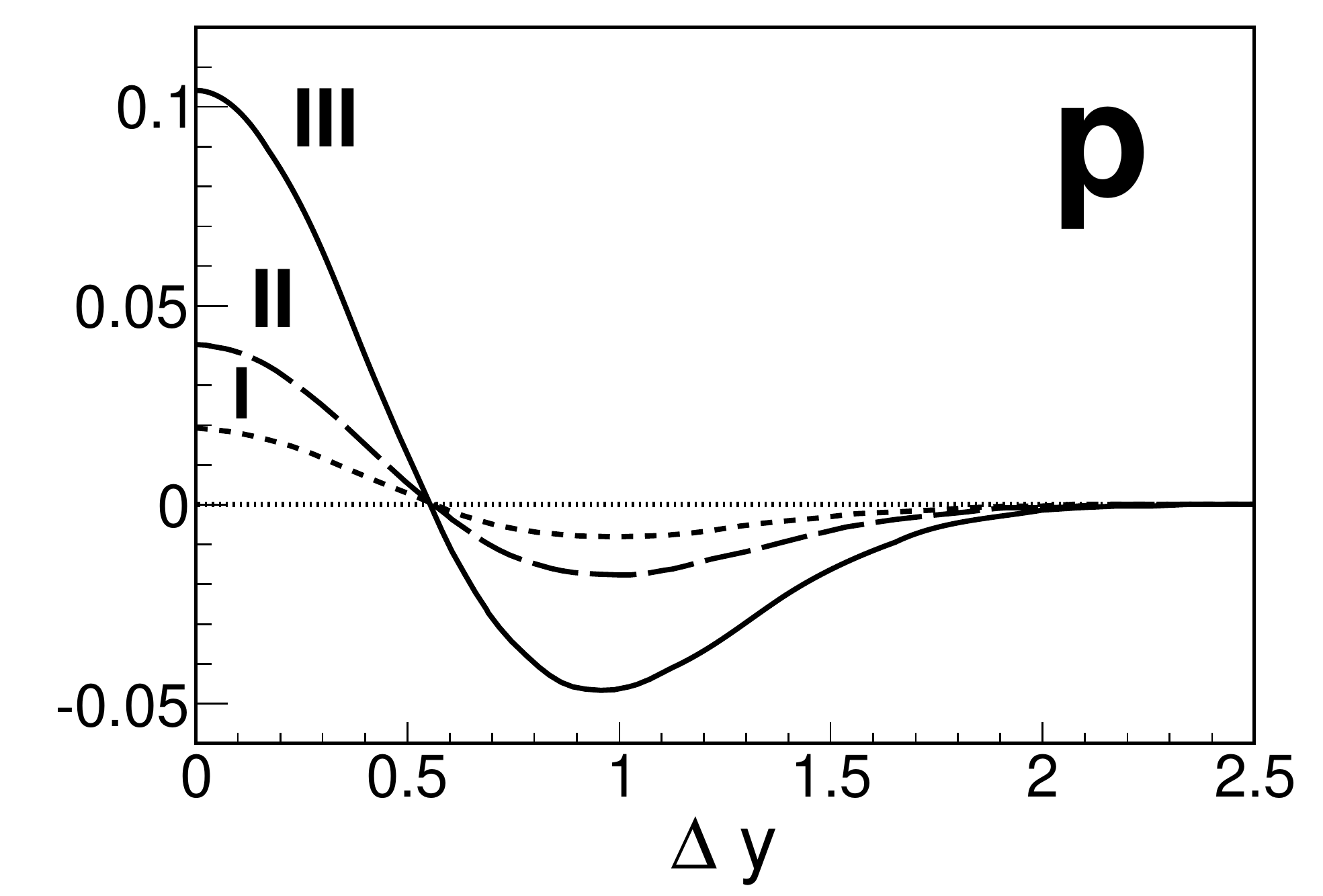}
\end{center}
\caption{Proton correlation function for the three trajectories.}
\label{fig:3in1_proton}
\end{figure*}
there is an upwards fluctuation $\delta \mu$ some point in space and time, that fluctuation will propagate outwards with the speed of sound.  If two points in space receive that signal at the same time, they will be positively correlated.  This is ultimately reflected in the spectrum of particles produced by the breakup of the fluid elements at the freezeout time via the fluctuations in temperature, chemical potential, and flow velocity in the Cooper-Frye formula.  Not surprisingly the biggest effect of the thermal conductivity is on fluctuations in the baryon chemical potential.  The effect on the two proton correlation function in rapidity is shown in Fig. 3.  Trajectory III obviously has the biggest effect and would clearly be observable.

More details of our calculations may be found in \cite{JoeJuan}.\\

This work was supported by the U.S. DOE Grant No. DE-FG02-87ER40328 and grant FPA2011-27853-C02-01.  J. M. T.-R. was also supported by an FPU grant from the Spanish Ministry of Education.

%\section*{References}

\end{document}